\documentclass[journal]{IEEEtran}
\ifCLASSINFOpdf
\else
\fi

\usepackage{algorithm}  
\usepackage{algpseudocode}  
\usepackage{amsmath} 
\usepackage{amssymb}
\usepackage{array}
\usepackage{booktabs}
\usepackage{cite}
\usepackage{color,soul}
\usepackage{enumitem}
\usepackage[export]{adjustbox}
\usepackage{fancyhdr}
\usepackage{float}
\usepackage{graphicx}
\usepackage{hyperref}
\usepackage{lineno}
\usepackage{listings}
\usepackage{mdwmath}
\usepackage{mdwtab}
\usepackage{multirow}
\usepackage{multicol}
\usepackage{rotating}
\usepackage{setspace}
\usepackage{soul}
\usepackage{url}
\usepackage{verbatim}
\usepackage{xspace}
\usepackage{subfigure}
\usepackage{caption2}

\usepackage{hyperref}
\usepackage{url}

\soulregister\cite7
\soulregister\ref7

\usepackage[textsize=scriptsize,backgroundcolor=yellow!40]{todonotes}

\IEEEoverridecommandlockouts

\hypersetup{draft}

\begin{document}
%
\title{Blockchain-based Federated Learning for\\ Device Failure Detection in Industrial IoT}

\author{
\IEEEauthorblockN{
Weishan Zhang\IEEEauthorrefmark{1},
Qinghua Lu\IEEEauthorrefmark{2}\IEEEauthorrefmark{3}\thanks{Qinghua Lu is the corresponding author. Email: qinghua.lu@data61.csiro.au},
Qiuyu Yu\IEEEauthorrefmark{1},
Zhaotong Li\IEEEauthorrefmark{1},\\
Yue Liu\IEEEauthorrefmark{3}\IEEEauthorrefmark{2},
Sin Kit Lo\IEEEauthorrefmark{3}\IEEEauthorrefmark{2},
Shiping Chen\IEEEauthorrefmark{2}\IEEEauthorrefmark{3},
Xiwei Xu\IEEEauthorrefmark{2}\IEEEauthorrefmark{3},
Liming Zhu\IEEEauthorrefmark{2}\IEEEauthorrefmark{3}\\
}
\IEEEauthorblockA{\IEEEauthorrefmark{1}College of Computer Science and Technology, China University of Petroleum (East China), China}\\
\IEEEauthorblockA{\IEEEauthorrefmark{2}Data61, CSIRO, Australia}\\
\IEEEauthorblockA{\IEEEauthorrefmark{3}School of Computer Science and Engineering, University of New South Wales, Australia}
}

\maketitle

\begin{abstract}

Device failure detection is one of most essential problems in industrial internet of things (IIoT). However, in conventional IIoT device failure detection, client devices need to upload raw data to the central server for model training, which might lead to disclosure of sensitive business data. Therefore, in this paper, to ensure client data privacy, we propose a blockchain-based federated learning approach for device failure detection in IIoT. First,  we  present  a  platform architecture  of  blockchain-based  federated  learning  systems  for failure detection in IIoT, which enables verifiable integrity of client data. In the architecture, each client  periodically creates  a  Merkle  tree  in  which  each  leaf  node  represents  a client  data  record,  and  stores  the  tree  root  on  a  blockchain. Further, to address the data heterogeneity issue in  IIoT  failure  detection,  we  propose  a  novel  centroid  distance weighted  federated  averaging  (CDW\_FedAvg)  algorithm  taking into  account  the  distance  between  positive  class  and  negative class  of  each  client  dataset. In addition, to motivate clients to participate in federated learning, a smart contact based incentive  mechanism  is  designed  depending on the  size and  the  centroid  distance  of  client  data  used  in  local  model training. A prototype of the proposed architecture is implemented with  our  industry  partner,  and  evaluated  in  terms  of  feasibility, accuracy  and  performance.  The  results  show  that  the  approach is feasible,  and  has  satisfactory  accuracy and performance.

\end{abstract}


%
\IEEEpeerreviewmaketitle

\section{Introduction}


\IEEEPARstart{D}{evice} failure detection is one of the most essential problems in industrial internet of things (IIoT)~\cite{9142419,IoTCop}. In conventional device failure detection of IIoT, client devices need to upload local raw data to the central server for centralised model training. This leads to the issue of data privacy as clients' local data might be business sensitive. For example, the hotels' air conditioning usage data might reflect occupancy rate.

Federated learning is an emerging machine learning paradigm~\cite{DBLP:journals/corr/KonecnyMYRSB16, 2020SLR} in a way that preserves privacy and reduces bias in model training. In each round of federated learning, multiple clients (e.g. organizations, data centers, IoT or mobile devices) are selected to train models locally to produce a global model while raw data are stored in clients and not exchanged or transferred.

Blockchain has been recently leveraged in IIoT federated learning to provide data integrity and incentives to attract sufficient client data and computation resources for training \cite{8733825,8894364,8892848}. However, there is a lack of systematic and holistic architecture design to support methodical development and efficient methods to tackle the challenge of data heterogeneity in device failure detection of IIoT. 

Therefore, in this paper, we present a blockchain-based federated learning approach for device failure detection in IIoT. To produce an unbiased global model, the model updates are aggregated using a novel averaging algorithm called centroid distance weighted federated averaging (CDW\_FedAvg), which takes into account the distance between the positive class and the negative class of each client dataset in weight calculation. In this study, ``positive class" means the detected failures, while ``negative class" indicates the normal operations of IoT devices. Client data are hashed and stored periodically on blockchain to ensure data integrity while addressing the performance issue of blockchain. Incentives are calculated based on clients' data contributions (i.e., the size and centroid distance of client data) and rewarded to motivate the clients via a smart contract. A proof-of-concept prototype of the proposed architecture is implemented with our industry partner and evaluated in terms of feasibility, accuracy, and performance. The evaluation results show that the proposed architecture is feasible and has satisfactory detection accuracy and performance.

The contributions of the paper are as follows.
\begin{itemize}
  \item A platform architecture that provides a systematic view of system interactions and serves as a guidance for the design of the blockchain-based federated learning systems in IIoT failure detection. The architecture design involves the following architectural design decisions: placement of model training, storage of monitored client data, incentive mechanism for clients, and deployment of blockchain. 
  \item A federated averaging algorithm called centroid distance weighted federated averaging (CDW\_FedAvg) taking into account the distance between the positive class and the negative class of each client dataset to reduce the impact of the data heterogeneity issue in IIoT device failure detection.
  \item A blockchain anchoring protocol which builds a custom Merkle tree in which each leaf node represents a record of data collected by a client's edge device, and places the root of the custom Merkle tree at a pre-configured interval on blockchain to ensure verifiable integrity of client data in an efficient way.
  \item An incentive mechanism via the designed incentive smart contract to motivate clients to contribute their data and computational resources for model training. The tokens rewarded to each client is dependent on the size and the centroid distance between the positive class and the negative class of data contributed to model training.
  \item Feasibility, detection accuracy and performance evaluation using a real industry use case which detects failures of water-cooled magnetic levitation chillers in air-conditioners.
\end{itemize}

The remainder of this paper is organized as follow. Section \ref{background} summarises the related works. Section \ref{architectureSection} presents the proposed solutions. Section \ref{evaluation} discusses the implementation and evaluation. Section \ref{conclusion} concludes the paper and outlines the future work.

\section{Related Work}
\label{background}

IoT has been integrated into industrial systems to allow a higher degree of automation and improve the economic benefits including productivity and efficiency. In industrial IoT, traditional centralized cloud-based computing approach may be inefficient, since tasks are pushed to cloud from numerous IoT devices which may cause data leakage and network overload \cite{7469991}. To address the issue of data privacy and limited bandwidth in cloud computing, edge computing has been widely applied in IIoT systems to shift part of computing tasks from cloud servers to edge nodes, which are closer to where data are generated. 

The concept of federated learning is first proposed by McMahan et al.~\cite{DBLP:journals/corr/McMahanMRA16}, which enables edge computing from the machine learning perspective. In a border definition of federated learning, multiple clients are coordinated to train model locally and aggregate model updates to achieve the learning objective, while raw data are stored in clients and not exchanged or transferred~\cite{kairouz2019advances, 2020SLR}. Compared with most studies of federated learning with central server, a few research works~\cite{zhao2019mobile, 8733825} build a federated learning framework without central server, such as applying p2p structure. 

Federated learning is classified into two types: cross-device federated learning and cross-silo federated learning. The former is introduced on mobile and edge device applications. For example, Google has applied federated learning setting to the prediction of search suggestions, next words and emojis, and the learning of out-of-vocabulary words~\cite{DBLP:journals/corr/abs-1812-02903, chen2019federated, ramaswamy2019federated}. The latter represents another type of federated learning setting in which clients are different organizations or geo-distributed datacenters~\cite{kairouz2019advances}. For example, Sheller et al. \cite{10.1007/978-3-030-11723-8_9} adopt cross-silo federated learning to build a segmentation model on the brain tumor data from multiple institutions. 

There have been dramatically surged numbers of studies conducted to address statistical heterogeneity of federated learning. Nishio et al.~\cite{8761315} design a novelty federated learning protocol with consideration of heterogeneous client properties, to reduce the impact of skewed data. Han et al.~\cite{aaai788} discuss to cope with the negative impact of the systematic data corruption  via collaborative machine teaching in FL. Pang et al.~\cite{9134408} propose a deep deterministic policy gradient based method for updates aggregation to deal with the issue of statistical heterogeneity in IoT. However, the effect of centroid distance of dataset is not considered in the above studies, which may cause influence to the overall performance.

Blockchain was first introduced as the technology supporting Bitcoin~\cite{Satoshi:bitcoin}, and has been generalized to an append-only data store which is distributed across computational nodes and structured as a linked list of blocks containing a set of transactions~\cite{8933021,9086020}. Currently, many studies have integrated blockchain into their federated learning approaches. Since federated learning relies on a single server, which is vulnerable to malfunction, some studies are focused on eliminating single-point failure in federated learning using blockchain~\cite{zhao2019mobile, 8733825}. Zhao et al.~\cite{zhao2019mobile} design a system based on federated learning and blockchain to let clients empower full nodes and compete to serve as a central server by turns for aggregation. Kim et al.~\cite{8733825} propose a blockchained-federated learning architecture in which participant clients are able to store local model updates on blocks and all clients as miners can access and aggregate the updates through smart contracts, without a single central server. Some works leverage blockchain in federated learning for auditability and data privacy \cite{8843900, 8892848}. Lu et al.~\cite{8843900} incorporate federated learning and blockchain into a data sharing scheme to preserve privacy. Majeed et al.~\cite{8892848} record local model parameters in each iteration into blocks, and introduces ``the global model state trie" to securely store global model. Incentive mechanism is another important research direction, which is imported to fedearted learning to encourage participants and standardize the behavior of participants~\cite{8894364, 8832210}. Weng et al. \cite{8894364} employ incentive mechanism and blockchain transactions to protect the privacy of local gradients, and enable auditability of the training process. Kang et al.~\cite{8832210} propose an incentive mechanism based on blockchain combing the reputation which is introduced to measure the trustworthiness and reliability of clients, to stimulate honest clients with high-quality data to participate in federated learning. Moreover, there are some recently proposed platforms which increase more novel participant roles, such as buyers who can pay for the clients in order to complete their training tasks \cite{8905038}. Bao et al.~\cite{8905038} provide a platform for buying federated learning models and design an architecture to reduce the time cost of buyers querying and verifying models on a blockchain.

However, the above studies do not present the architecture design details of blockchain-based federated learning systems with consideration of the data heterogeneity issue in IIoT failure detection. Additionally, how to ensure the integrity of client data is not discussed. In this paper, we propose an architecture of federated learning for IIoT failure detection using blockchain in a way that motivates clients to provide sufficient data and computing resources for model training while reducing the impact of data heterogeneity, and preserving client data privacy and integrity.

\begin{figure*}[!t]
	\centering
	\includegraphics[width=0.9\linewidth]{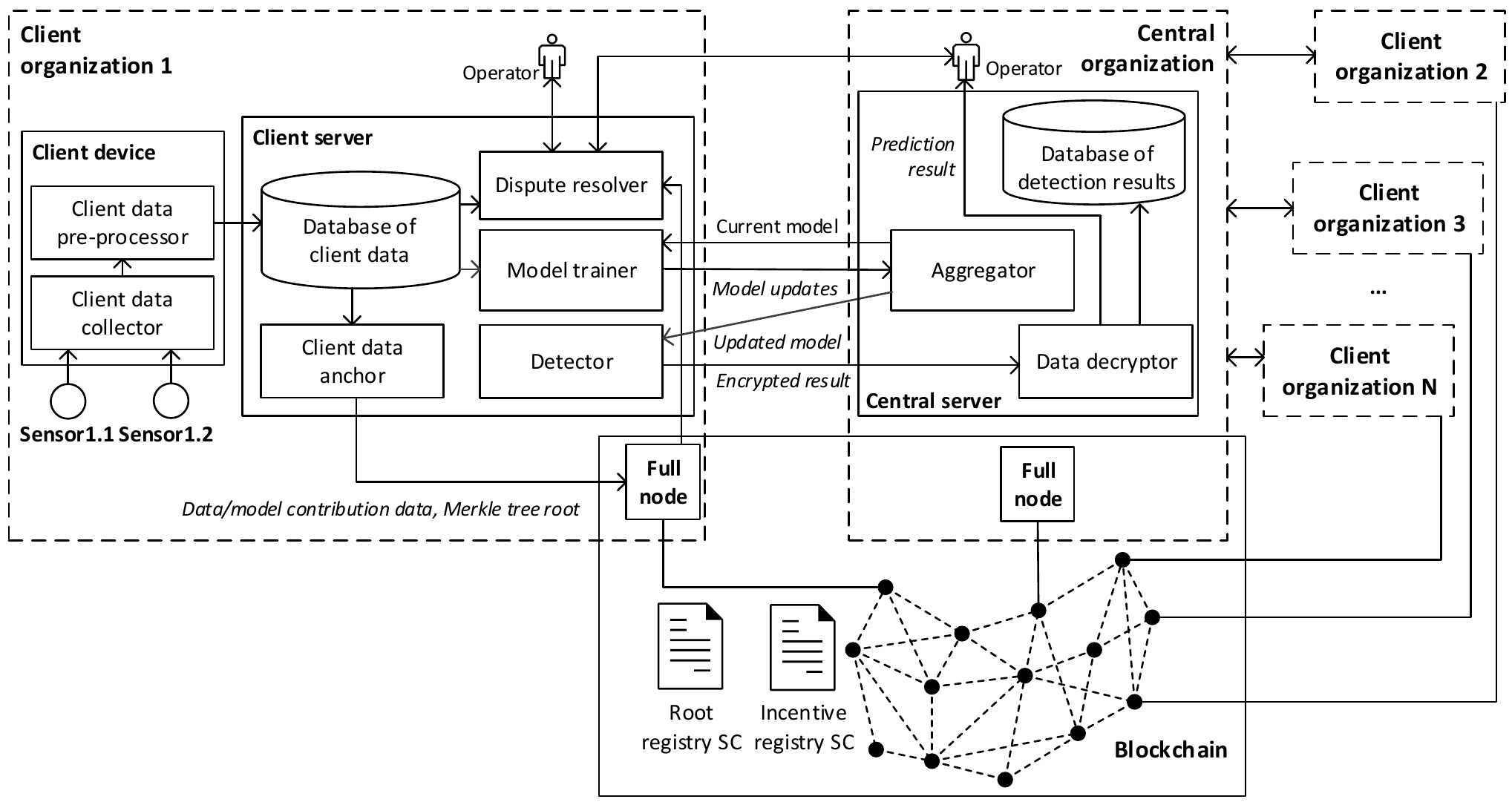}
	\caption{Architecture of blockchain-based federated learning systems for failure detection in IIoT.}
	\label{architecture}
\end{figure*}

\section{Approach}
\label{architectureSection}
We adopt the standard requirements elicitation methodology \cite{Kotonya:1998:REP:552009} for our industry partner's specific requirements and derive a list of general application-agnostic functional and non-functional requirements for blockchain-based federated learning systems for IIoT device failure detection. The gathered functional requirements include: 1) Making detection using a trained model; 2) Resolving disputes between platform owner and clients; 3) Rewarding clients for their contributions to the model training process. The identified non-functional requirements include: 1) Preserving edge data privacy; 2) Maintaining satisfactory detection accuracy; 3) Ensuring edge data integrity; 4) Keeping availability of blockchain infrastructure.

In this section, we present a platform architecture for blockchain-based federated learning systems to meet the above general application-independent requirements for federated learning systems. Section \ref{subsect:Overall Architecture} provides an overview of the architecture and discusses how the components and their interactions.  Section \ref{subsec:Architectural Design Decisions} discusses the main architectural decisions. Section \ref{subsec:weighted} proposes a novel federated averaging algorithm called centroid distance weighted federated averaging (CWD\_FedAvg). Section \ref{subsec:Anchoring protocol} presents a protocol for anchoring the client data to the blockchain to ensure data integrity. Section \ref{subsec:incentive} designs an incentive mechanism to encourage the clients to contribute to model training.

\subsection{Overall Architecture} \label{subsect:Overall Architecture}
Fig.~\ref{architecture} illustrates the overall platform architecture, which consists of two types of participating organizations: central organization and client organization. The central organization is the federated learning platform owner (e.g. manufacturer company) which maintains all industrial services based on detection results analyzed by the platform, while the client organizations are the organizations that contribute local data and computation resources for local model training.

Each client organization maintains sensors, a client device, a client server, and a blockchain full node, while the central organization hosts a central server and a blockchain full node. Each client device (e.g., Raspberry Pi) gathers the environment data detected by sensors through \emph{client data collector} and clean the data (e.g., noise reduction) via \emph{client data pre-processor}. All the collected client environment data are stored in \emph{database of client data} hosted by the client server. If a client organization is selected by the central server for a certain round of model training, the client server downloads the current version of the global model from the central server to further train the model and compute model updates using the local data via \emph{model trainer}. The model updates are sent to the \emph{aggregator} on the central server at a pre-configured round. \emph{Aggregator} combines the model updates received from the clients and produces a global model, which is then sent back to client servers for \emph{Detector}. The detection results are encrypted and sent to the central organization, where the operator can decrypt the received detection results via \emph{data decryptor} and make the final decision.

Each client server periodically creates a Merkle tree in which each leaf node represents a data record collected by sensors. The information of a client in each selection round (including  the  Merkle  tree  root, the status of training, the size and the centroid distance of client data for model training) are all stored in the pre-deployed smart contracts on the blockchain through \emph{client data anchor}. If a dispute (e.g. about failure cause) occurs, the operators of both the central organization and the client organization can use \emph{dispute resolver} to verify the integrity of client historical data in a certain anchoring period, by comparing the data with the respective Merkle tree root stored on-chain. To reward the client organizations for their contribution to model training, tokens are distributed to them according to the contributed size and centroid distance of client data used for model training.

\subsection{Architectural Design Decisions} \label{subsec:Architectural Design Decisions}
\subsubsection{Local Model Training}
To preserve the privacy of client data (e.g. usage frequency and time), which is often sensitive to each client organization's business, the architecture adopts federated learning to train the models locally on client servers via \emph{model trainer} while computing training results on the central server to formulate a global model for detection via \emph{aggregator}.

\subsubsection{Anchoring Monitored Data to Blockchain}
To resolve disputes between the central organization and client organizations (e.g. about failure causes), the architecture leverages the blockchain to enable verifiable integrity of client data via \emph{client data anchor} on the client server. Since blockchain has limited storage capabilities, the client server periodically creates a Merkle tree in which each leaf node represents a record of data collected by sensors, and stores its root in the pre-deployed \emph{root registry smart contract (SC)} through \emph{client data anchor}.

\subsubsection{Smart Contract based Incentive Mechanism}
The architecture provides incentives to encourage client organizations for model training. Tokens are rewarded to client organizations based on the size and the centroid distance of data taken in model training. The rewarded tokens are recorded in \emph{incentive registry SC}. 

\subsubsection{Blockchain Deployment}
In a blockchain network, a full node maintains a complete list of every single transaction that had occurred on the blockchain. In the proposed architecture, each participating organization, including central organization and client organization, hosts one blockchain full node. Hence, each organization has a full replica of the data stored on the blockchain which can be used for auditing, and ensure the availability of the whole system.

\begin{algorithm}[!h]
\caption{CWD\_FedAvg Algorithm}  
  \label{fedavg}
  \footnotesize
  \begin{algorithmic} 
    \State /*Central server*/
    \State Initialize $w_0$
    \For {each round t = 1, 2, ...}
         \State Select $K$ clients
         \For {each client $k \in K$ clients}
              \State $w^k_t \gets$ UpdateClient( )
              \State $d\_^k_t$ $\gets$ the distance between two classes in training dataset
         \EndFor
         \State $f(d^k_t) \gets \frac{1}{d^k_t}$
         \State $w_t \gets \sum_{k=1}^K n_k \ast f(d^k_t) \ast w_{t-1} / \sum_{k=1}^K n_k \ast f(d^k_t) $
    \EndFor
    \State /*Client update*/
    \State UpdateClient( )\{
    \State Initialize local minibatch size $B$, local epochs $E$, learning rate $\eta$
     \For{each epoch $i \in E$ }
          \State randomly sample $S_i$ based on size $B$
          \State $w_i$ $\gets$ $w_{i-1}$ - $\eta\bigtriangledown$g($w_{i-1}$ ; $S_i$ )
     \EndFor
     \State return $w_i$
     \State \}
  \end{algorithmic}
\end{algorithm}

\subsection{Centroid Distance Weighted Federated Averaging} \label{subsec:weighted}
The federated averaging (FedAvg) algorithm proposed by Google~\cite{DBLP:journals/corr/KonecnyMYRSB16} is not suitable for industrial IoT systems since the heterogeneity between the different client local datasets exists heavily in industrial IoT systems (i.e., the distribution and size of the dataset on each client may be different). Take the air-conditioner failure detection as an example, the deployed air-conditioners might have different working environments (e.g., weather) and different users (e.g., age).

To reduce the impact of data heterogeneity and improve the model performance, we propose a centroid distance weighted federated averaging (CDW\_FedAvg) algorithm taking into account the distance between positive class and negative class of each client dataset. The weighted average process are calculated as:

\begin{equation}
\label{weightedavg}
w_t = \frac{\sum_{k=1}^K s^k_t\ast f(d^k_t)\ast w_t}{\sum_{k=1}^K s^k_t\ast f(d^k_t))},
h(d^k_t) = 1/d^k_t
\end{equation}
where \emph{$s^k_t$} represents the size of \emph{k\_th} client training dataset contributed in \emph{t\_th} round. \emph{$d^k_t$} denotes the distance between the centroid of positive class and the centroid of negative class from the \emph{k\_th} client dataset in \emph{t\_th} round. \emph{$f(d^k_t)$} is the function to process \emph{$d^k_t$}, outputting the reciprocal of \emph{$d^k_t$}. Here, we sacrifice the model accuracy for client datasets which have greater distance between the centroid of positive class and centroid of negative class since most of the client datasets in the real-world have smaller distance. The detailed process is illustrated in Algorithm~\ref{fedavg}.

In Algorithm~\ref{fedavg}, for each round, the central server selects a subset of clients to participate in the round based on some predefined criteria (e.g., idle and charging) and sets the local minibatch size \emph{B}, local epochs \emph{E} and learning rate $\eta$ on the selected clients. Then, the selected clients execute local Stochastic Gradient Descent (SGD) on \emph{B} batches in \emph{E} epochs, calculate the distance \emph{$d^k_t$} and return the update \emph{$w_t$} and \emph{$d^k_t$} to the central server. Finally, on the central server, the updates are weighted averaged according to the size of local training dataset and the reciprocal of \emph{$d^k_t$}.

\begin{algorithm}[!h]  
  \caption{Anchoring Protocol}  
  \label{anchoring_protocol}
  \footnotesize
  \begin{algorithmic}[1]
  \For{each period $p_i$ = 1, 2, ...}
   \For{$c_j \in C$}
      \For{$d^c_{pn}$}
          \State $d^c_{pn}$ $\gets$ String( $d^c_{pn}$)
          \State $d^c_{pn}$ $\gets$ Hash( $d^c_{pn}$)
          \State $pNodeList[0..N] \gets$ $d^c_{pn}$
       \EndFor
       \Repeat
        \For{ ($k$ = 0 ; k $\textless$ length($pNodeList$) ; k+2)}
           \State $l \gets pNodeList[k]$
           \If{$k$ = length($pNodeList$)}
               \State $r \gets l$
           \Else
                \State $r \gets pNodeList[k + 1]$ 
           \EndIf
           \State $r \gets$ SHA256($l+r$)
           \State $temp[j] \gets r $ 
           \State (each $j$ $\in$ [0,$\lceil$k/2$\rceil$])
        \EndFor
        \State $pNodeList[0..j] \gets temp[0..j]$  
      \Until{length($pNodeList$) = 1}
      
      \State $root \gets$ $pNodeList[0]$
      \State $time \gets$ current time
      \State UpdRootSC($time, root$)
    \EndFor
    \EndFor
   \For{each training round }
     \State  model training
       \State UpdStaSC($rNo$, $dataSize$, $distance$)
     \State CalIncentiveSC( )
   \EndFor
  \end{algorithmic}  
\end{algorithm}

\subsection{Anchoring Protocol} \label{subsec:Anchoring protocol}
An anchoring protocol is designed to build a Merkle tree based on the monitored data collected by a client, and anchor the Merkel tree root to the blockchain for client data integrity and auditing (i.e., dispute resolution between the central organization and the client organization). The incentive mechanism is also supported by the anchoring protocol to reward the client organizations according to the anchored contribution recorded on the \emph{incentive registry SC}.

Algorithm~\ref{anchoring_protocol} describes how the anchoring protocol works. The anchoring protocol is scheduled according to the agreement between the central organization and client organizations (e.g. every hour). For each anchoring period $p_i$, the protocol goes through each client organization $c_j$, and queries the client data $d^c_{pn}$ collected within the anchoring period. The protocol converts each data record $d^c_{pn}$ to a string format. All the data collected within the anchoring period are hashed and used to construct a Merkle tree where each leaf node represents the hash value of a data record. 

To construct the Merkle tree structure, the protocol first creates a \emph{pNodeList} to store the parent node values and a \emph{temp} to put the results of the combined hash values of the two child nodes. All the leaf node values are initially placed in the \emph{pNodeList}. To compute the parent node of each pair of two leaf nodes which are adjacent to each other, the hash values of them are assigned to \emph{l} which denotes left child node, and \emph{r} which denotes right child node respectively, and combined to produce the hash value for the parent node, which is then added to \emph{temp}. After going through all the elements in \emph{pNodeList}, the \emph{temp} is converted to \emph{pNodeList}. The tree construction process is executed recursively until the length of the \emph{pNodeList} becomes one. The root is exactly the only value in the \emph{pNodeList}.

\emph{UpdRootSC()} is used to anchor the Merkle tree \emph{root} and timestamp of the current period to \emph{root registry SC}. Once the client servers finish their work and sent the model updates to the central server, \emph{dataSize} (the size of data for training) and \emph{distance} (the centroid distance between two classes of training dataset) of each participant client server and the serial number of training round \emph{rNo}  is anchored in the blockchain by \emph{UpdStaSC()}. The corresponding tokens are then calculated and rewarded to client organizations through \emph{CalIncentiveSC()}. Both \emph{UpdStaSC()} and \emph{CalIncentiveSC()} are encoded in Incentive registry SC.

\subsection{Incentive Mechanism} \label{subsec:incentive}
Algorithm~\ref{incentive} describes how the incentive mechanism works. The incentive mechanism aims to reward client servers with tokens, according to their contribution (the size of data applied for model training and the distance between two classes of training dataset). For every training round, the central server chooses a fixed number of client servers, and stores their blockchain addresses in \emph{Clist}. Besides, the central server should also define a struct \emph{training} to enroll finished training tasks of client servers, including the status of training work \emph{finished}, the size of data for training \emph{dataSize}, and the centroid distance between two classes of training dataset \emph{distance}. The chosen client servers train models locally, and upload corresponding information into \emph{training}, which is then converted to \emph{contri}. Finally, the incentive \emph{tokens} are distributed to each \emph{address} of the client server that \emph{Clist} records, in accordance with their \emph{contri} respectively. The number of tokens is the sum of the \emph{dataSize} and a value computed by multiplying \emph{distance} by a constant \emph{C}. The ultimate goal of this incentive mechanism is to encourage more client servers for participation, to eliminate the model bias.

\begin{algorithm}[!h]    
\caption{Incentive Calculation and Distribution}  
  \label{incentive}
  \footnotesize
  \begin{algorithmic}[1]
  \State /*Central server executes*/
  \State $Clist$ $\gets$  selected clients address
  \State /*Client server executes*/
  \State $struct$ $\{$$finished, dataSize, distance$$\}$ $training$
  \For{each training round }
  \If{$address \in Clist$}
  \State model traning
  \State UpdStaSC($rNo$, $dataSize$, $distance$)
  \State CalIncentiveSC( )
  \EndIf
  \EndFor
  \State /*Update Status of Training work */

  \State UpdStaSC($rNo$, $dataSize$, $distance$)$\{$
    \State \hspace*{1em}  $training.finished \gets$ true
    \State \hspace*{1em}  $training.dataSize \gets dataSize$
    \State \hspace*{1em}  $training.distance \gets distance$
    \State \hspace*{1em}  $contri[addr][rNo] \gets work$
  \State $\}$
  \State /*Calculation Incentive*/
  \State CalIncentiveSC( )$\{$
    \hspace*{0.6cm}  \If{$address \in Clist$}
    \If{$contri[addr][rNo].finished$ == true}
      \State $token[addr]$ = $token[addr]$ + $contri[addr][rNo].dataSize$
      \State \hspace*{5.5em}  + $contri[addr][rNo].distance$ * $C$
      \EndIf
  \EndIf
  \State $\}$
  \end{algorithmic}  
\end{algorithm}

\section{Implementation and Evaluation}
\label{evaluation}
In this section, we implement a proof-of-concept prototype using a real-world use case, and evaluate the prototype in terms of feasibility accuracy, and performance.

\subsection{Use Case}
Our industry partner is one of the largest air-conditioner manufacturers in the world, and a large portion of their clients are hotels which are distributed across the globe. The failures of the hotel air-conditioning systems directly influence customer satisfaction. In the collaborated project, we aim to predict failures for air-conditioning systems. In the use case, we claim that the manufacturer owns the servers/devices placed in the hotels. The ownership of hotel servers/devices here means only the manufacturer has the write-access authority to the firmware/software on the devices/servers.

In particular, the project seeks to place most of the machine learning tasks within the hotels to preserve the data privacy of hotels. For example, room occupancy percentage is sensitive to a hotel and can be calculated based on the air-conditioning system data. Also, the project aims to encourage hotels to contribute their air-conditioner data and train failure detection models locally. Tokens are rewarded to hotels based on their contribution which can later be used to buy new products or services.

The water-cooled magnetic levitation chiller is the key component in a hotel air-conditioning systems. We developed a proof-of-concept prototype for the failure detection system of water-cooled magnetic levitation chillers using the proposed architecture.

\subsection{Implementation}
\begin{figure}[!t]
	\centering
	\includegraphics[width=0.9\columnwidth]{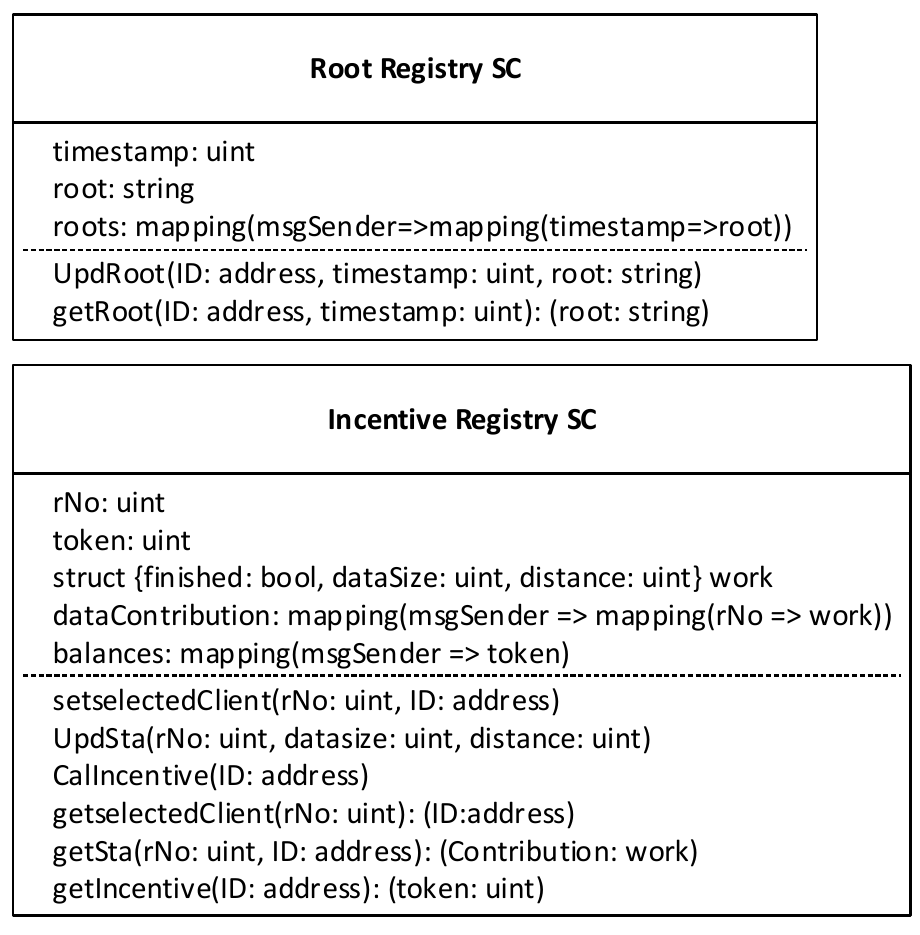}
	\caption{On-chain data structure}
	\label{on_chain_data_structure}
\end{figure}

Fig. \ref{on_chain_data_structure} shows the design of two on-chain smart contracts for client data anchoring and model training incentives: \emph{root registry SC} stores Merkle tree roots and corresponding timestamps, while \emph{incentive registry SC} maintains incentive information including training round number, the status of training task, size of client data, distance between two classes of client data applied for training work, and rewarded tokens based on data size and distance.

We adopt a federated learning framework named \emph{Leaf}~\cite{caldas2019leaf} in the implementation. We use Ethereum \cite{buterin2013ethereum} as the underlying blockchain network, in which the consensus algorithm is Proof-of-Work (PoW), and the difficulty is configured to $0x4000$. The smart contracts are written in Solidity and compiled with Solidity compiler version 0.4.20. The \emph{client data anchor} component is developed in Java 1.8. We select MySQL 5.7.25 as the supported database to store the off-chain data.

We deployed the implemented prototype on 5 Alibaba Cloud\footnote{\url{https://www.aliyun.com/}} virtual machines (2vCPUs, 8G RAM) as client/central servers (1 cloud server used as the central server and 4 cloud servers used as client servers) and 4 Raspberry Pi 3 boards configured to Raspbian 2018-11-3 as 4 client devices. Each client server is allocated to one hotel where installed one large air conditioner (as client device). The data collected by the sensors in each air conditioner includes 70 parameters. After reducing the noisy data and removing the redundant parameters (from two compressors), 18 parameters are kept for machine learning as client data, which are listed as follows.

\begin{itemize}[noitemsep,topsep=0pt]
    \item Evaporator inlet water temperature
    \item Evaporator outlet water temperature
    \item Condenser inlet water temperature
    \item Condenser outlet water temperature
    \item Evaporator cooling capacity
    \item Compressor inlet air temperature
    \item Compressor outlet air temperature
    \item Evaporator inlet air pressure
    \item Condenser outlet air pressure
    \item Exhaust air overheat temperature
    \item Main circuit's coolant level
    \item Opening size of the main coolant pipe valve
    \item Compressor's load
    \item Compressor's current
    \item Compressor's rotational speed
    \item Compressor's voltage
    \item Compressor's power
    \item Compressor's inverter's temperature

\end{itemize}

For the experiments, each data record consists of 18 features as mentioned above and 1 label which indicates whether any fault occurred. A client server has one classifier, which is a four-layer neural network. The first layer consists of 18 units corresponding to the 18 features. Then there are two hidden layers which both have 150 units. The output layer is a \emph{Softmax} function to classify normal data and abnormal data. Please note that the hidden layers apply \emph{ReLu} function as the activation function. The learning rate is set to 0.005.

\subsection{Feasibility}
\begin{figure}[!t]
	\centering
	\includegraphics[width=\columnwidth]{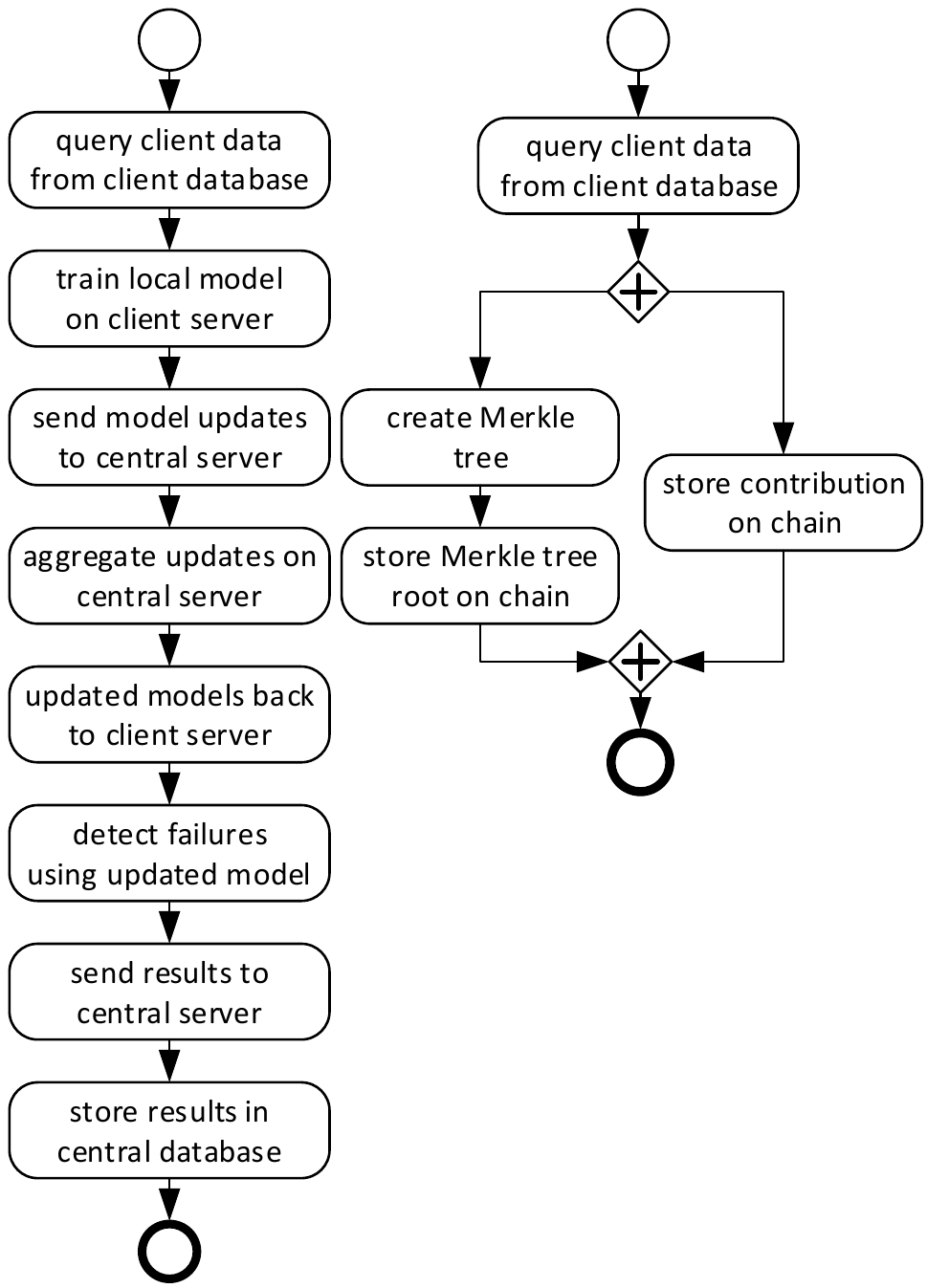}
	\caption{Federated learning process and blockchain anchoring process.}
	\label{process}
\end{figure}

To evaluate the feasibility of the proposed approach, we examined the implemented proof-of-concept of proposed architecture against the functional and non-functional requirements identified in Section III. Fig.~\ref{process} illustrates the workflow of federated learning and blockchian anchoring.

For the functional requirements, 1) the platform prototype is able to train local models on the client  server hosted by each hotel and aggregate the local model updates into a global model for failure detection. 2) The operators of the hotels and the air-conditioner platform are able to verify the integrity of client data for a certain anchoring period by comparing with the Merkle tree root stored on blockchain. 3) In addition, the hotels can be allocated with tokens as the reward for their contribution via \emph{Incentive Registry SC}.

Regarding the non-functional requirements, 1) the air-conditioner data privacy is preserved by the federated learning setting, in which the central server only needs air-conditioner model updates rather than all air-conditioner usage information. 2) In particular, the applied federated learning is more advantageous than centralized model training on the bandwidth cost, if the size of training data is large. 3) The detection accuracy is maintained by a shared model, which is formulated by aggregating updates from models trained on each hotel (client server). 4) Air-conditioner data integrity is achieved via the anchoring protocol. When data auditing is needed, the data stored locally in each hotel can be again constructed to the Merkle tree structure and verified by comparing it with the root stored on the blockchain. Each hotel and the air-conditioner manufacturer maintains a full node to keep the blockchain infrastructure available, in the case that some nodes maybe out-of-service.

\subsection{Quantitative Analysis - Accuracy and Performance}

In the quantitative analysis, each client server stores about 1,000 training records and 1,000 testing records. The training records are distributed on client servers, while a global model is learnt on the central server by aggregating local model updates. We first tested accuracy of the global model of failure detection on each individual client server with their own testing records. Moreover, we compared the accuracy of the global model trained using federated learning with the respective model trained using the traditional centralized learning and local learning. We tested two classification models including \emph{Logistic Regression (LR)} and \emph{Neural Networks (NN)}.

\begin{figure}[!t]
	\centering
	\includegraphics[width=\columnwidth]{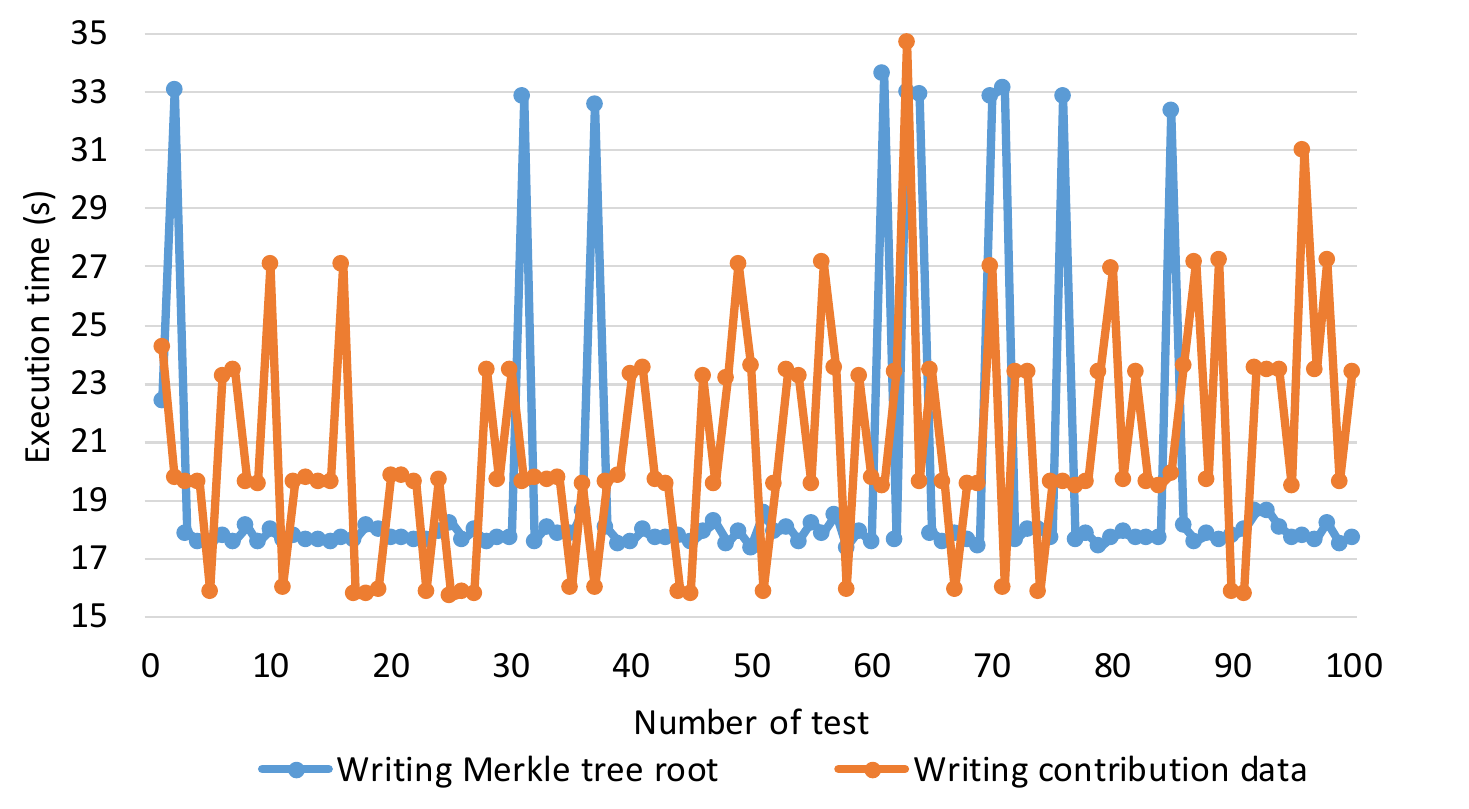}
	\caption{Blockchain execution time.}
	\label{blockchain_tx_time}
\end{figure}

\textit{Test 1: Measuring blockchain anchoring time}\par
The executing time of anchoring protocol includes querying data from the database, creating a Merkle tree, and writing the Merkle tree root and contribution data on the blockchain.
Fig.~\ref{blockchain_tx_time} only illustrates the execution time for writing Merkle tree root and contribution data to blockchain, since it costs most of the execution time for the anchoring protocol. We measured the transaction inclusion time which is for the transaction to be included in a block. 
Writing the Merkle tree root takes around 18s, while the average time for writing contribution data is around 21s. The writing latency on the blockchain is much longer than in the conventional database as it takes time to propagate transactions/blocks and achieve consensus. Since blockchain is used for auditing and incentives provision, the writing latency of blockchain is not a concern in this work. We have also examined the Merkle tree creation time, which takes about 250ms. As the Merkle tree creation time is much less than transaction inclusion time, it is not included in Fig.~\ref{blockchain_tx_time}.

\begin{figure}[!t]
	\centering
	\includegraphics[width=\columnwidth]{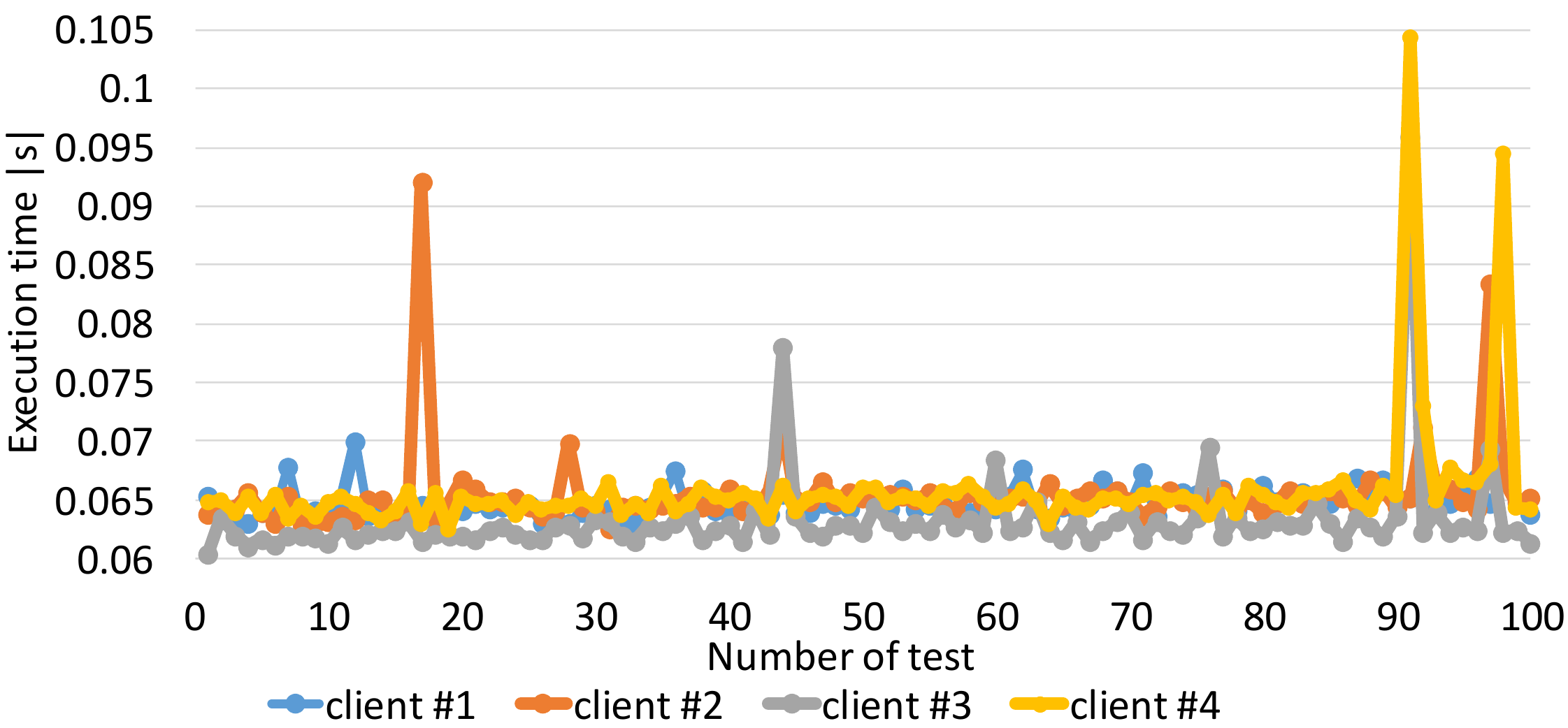}
	\caption{Execution time of incentive mechanism.}
	\label{incen_mech_time}
\end{figure}

\begin{figure}[!t]
	\centering
	\includegraphics[width=\columnwidth]{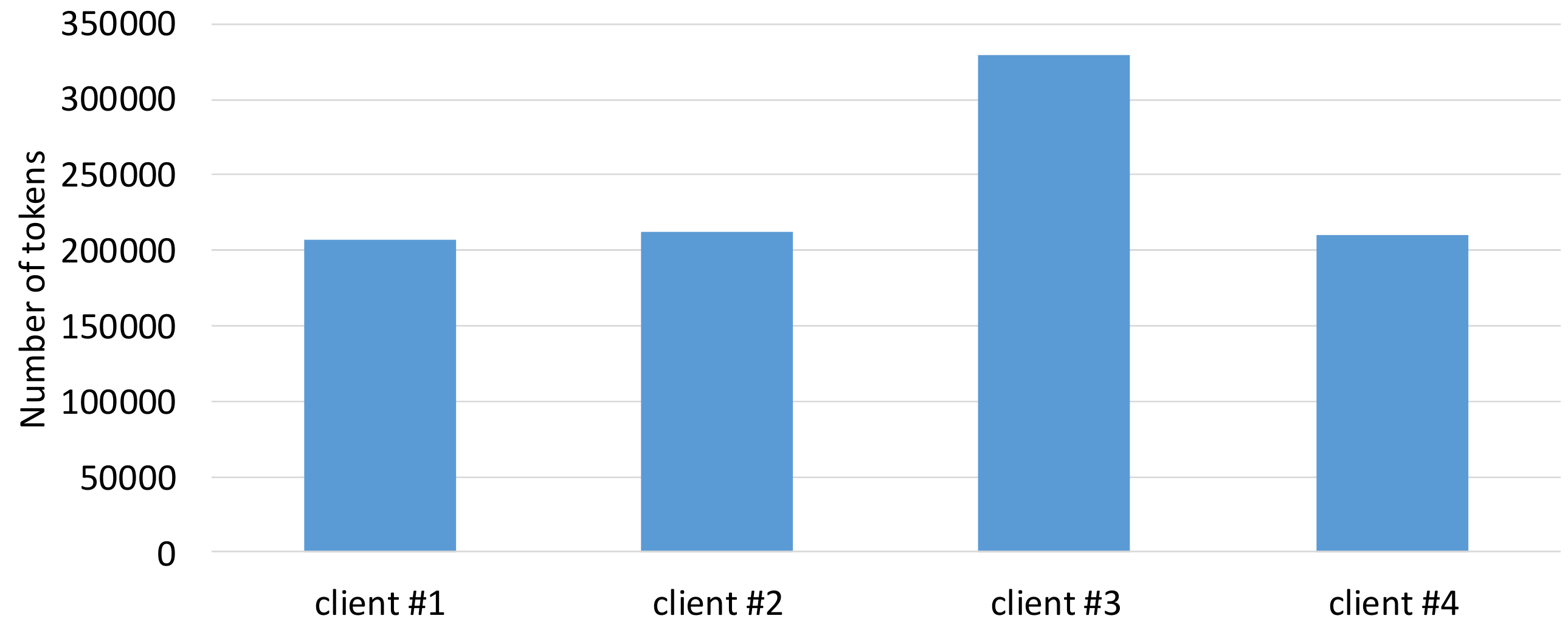}
	\caption{Tokens given to each client.}
	\label{incen_token}
\end{figure}

\begin{figure}[!t]
	\centering
	\includegraphics[width=\columnwidth]{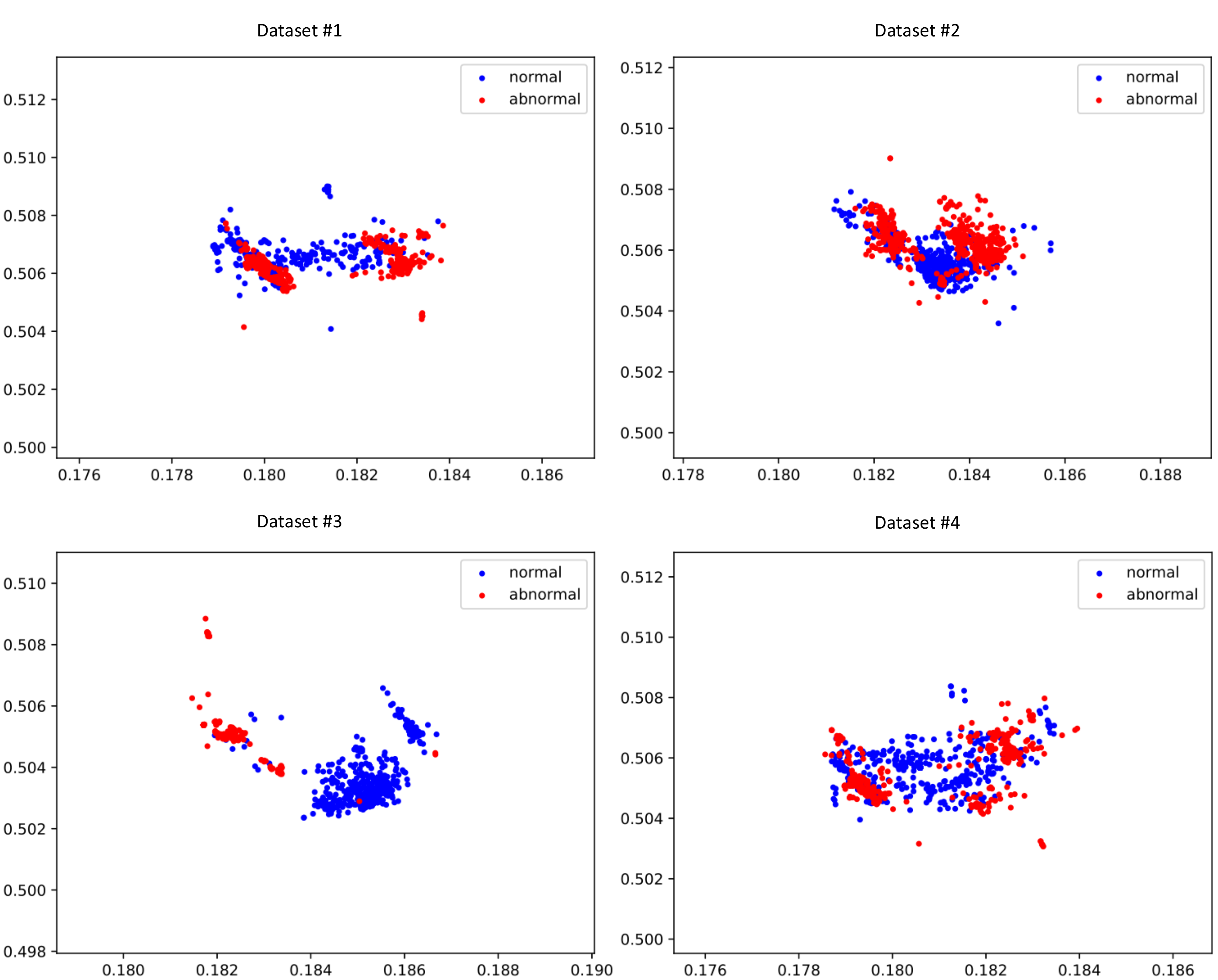}
	\caption{Training data distribution after dimension reduction.}
	\label{data_distribution}
\end{figure}

\textit{Test 2: Measuring incentive mechanism}\par
The executing time of incentive mechanism includes the time spent on calculating tokens rewarded  and writing them on blockchain. We measured the executing time of incentive mechanism. As shown in Fig.~\ref{incen_mech_time}, the average execution time is around 0.065s for all the four clients, which is efficient. Fig.~\ref{incen_token} presents the tokens rewarded to each of the four clients with different data distance. Client \#3 gains the largest number of tokens since it has the largest data distance. The results show that the incentive mechanism encourages clients to provide high quality training data.

\begin{figure*}[!ht]
	\centering
	\includegraphics[width=1\linewidth]{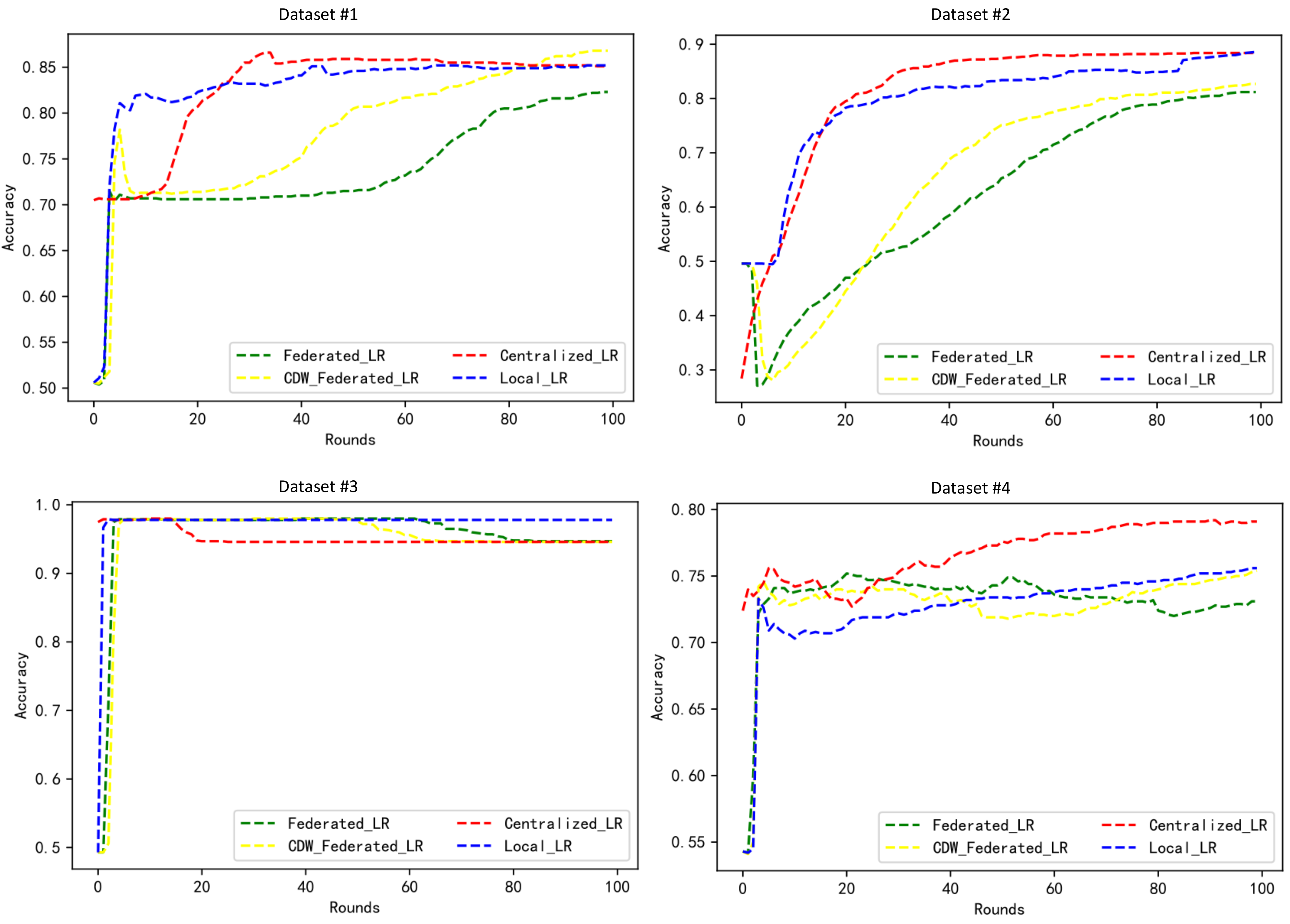}
	\caption{The accuracy of LR model.}
	\label{lr-accuracy}
\end{figure*}

\begin{figure*}[!ht]
	\centering
	\includegraphics[width=1\linewidth]{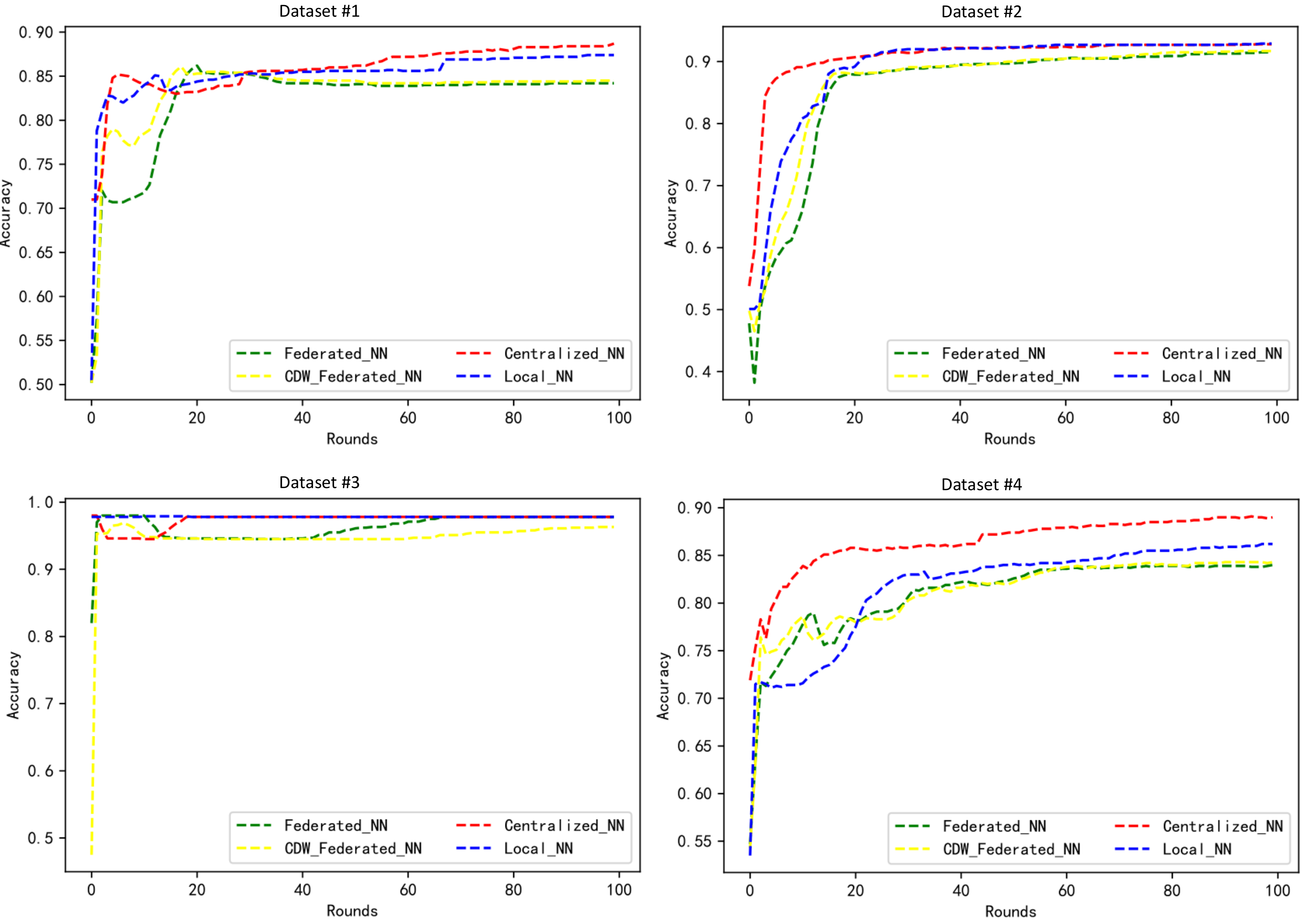}
	\caption{The accuracy of NN model.}
	\label{nn-accuracy}
\end{figure*}

\begin{figure*}[!ht]
	\centering
	\includegraphics[width=1\linewidth]{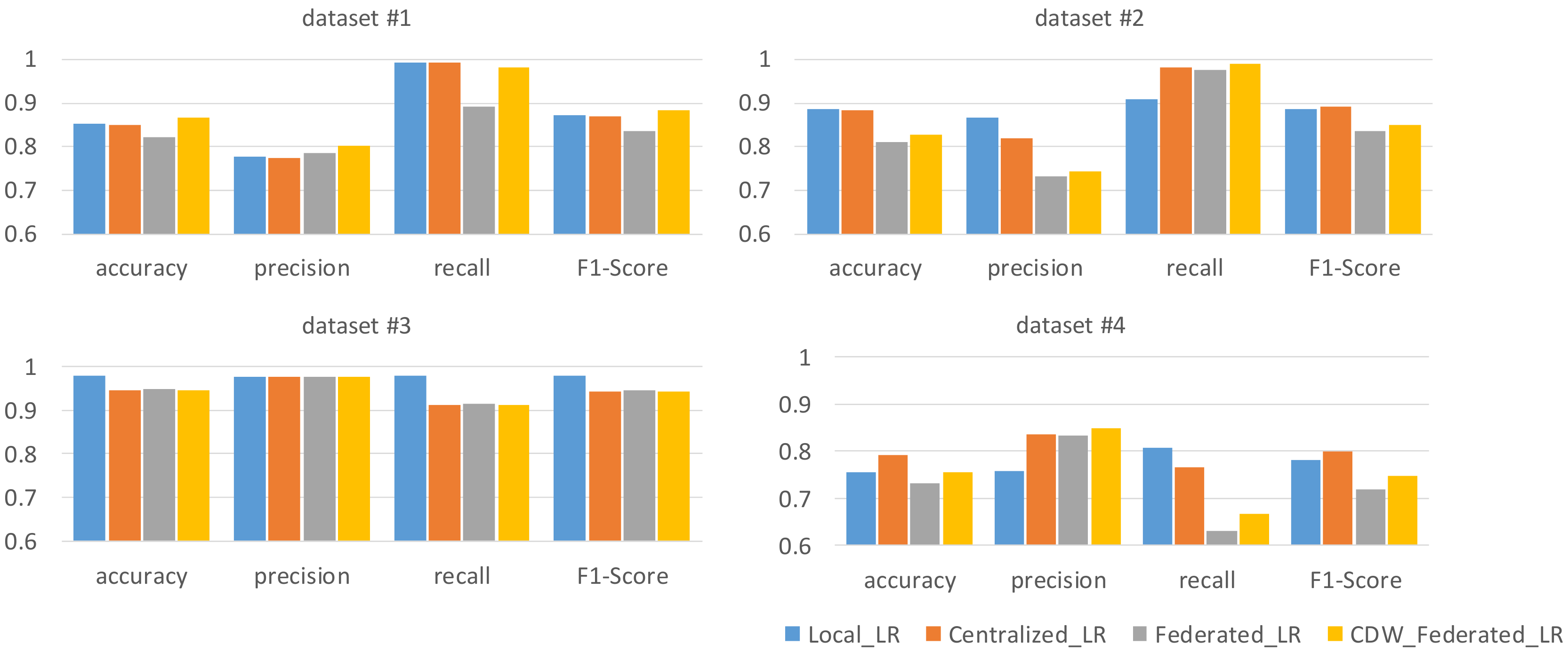}
	\caption{Precision, recall, and F1-score of LR model.}
	\label{lr-metrics}
\end{figure*}

\begin{figure*}[!ht]
	\centering
	\includegraphics[width=1\linewidth]{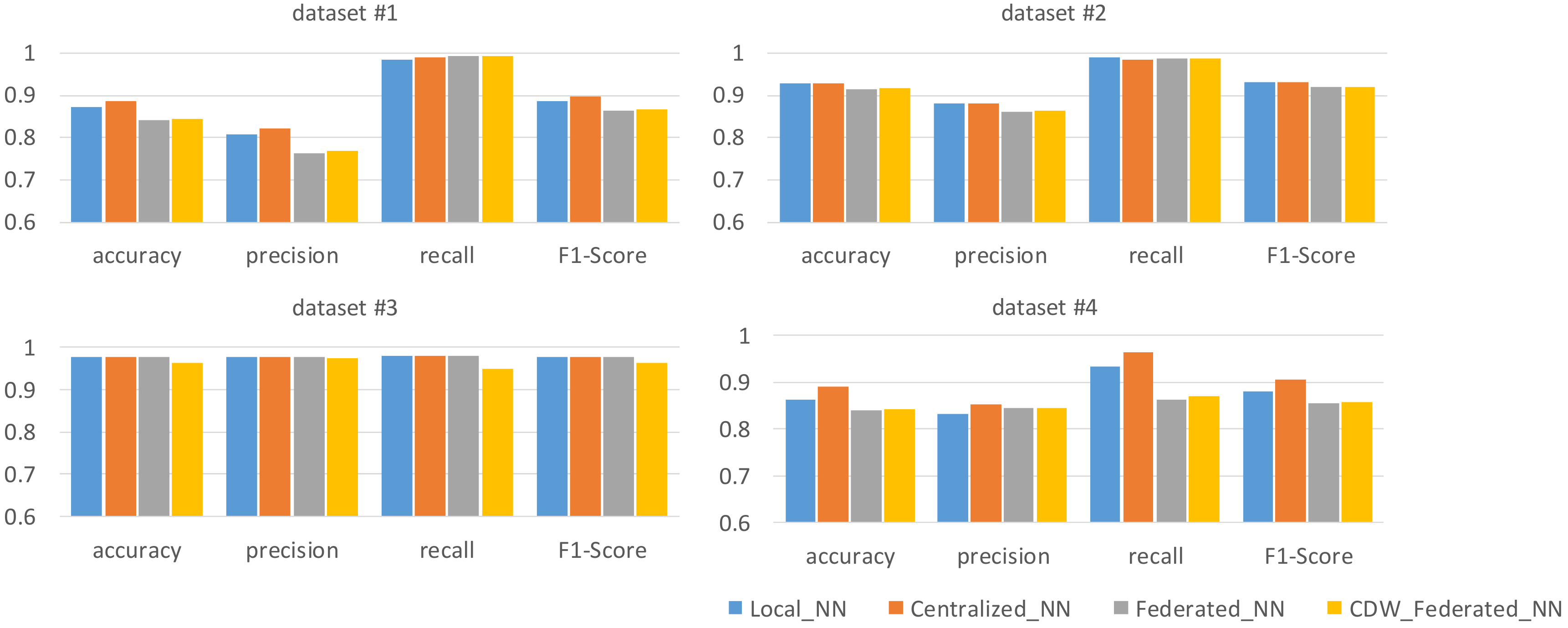}
	\caption{Precision, recall, and F1-score of NN model.}
	\label{nn-metrics}
\end{figure*}

\textit{Test 3: Measuring model detection accuracy}\par
Before we measured the model prediction, we first examined the distribution of each client dataset to check their distribution difference. Fig.~\ref{data_distribution} shows the distribution of positive class and negative class in each client dataset. We observed that there is a distance from positive class to negative class in client \#3, while other clients show a different pattern.

We measured the accuracy of \emph{Logistic Regression (LR)} and \emph{Neural Networks (NN)}, using proposed centroid distance weighted (CDW) federated learning method with CDW\_FedAvg, standard federated learning with FedAvg, centralized learning and local learning respectively for 100 training rounds. For each round of federated learning, the model is trained locally on each client through 40 epochs and the model updates are sent to central server to produce a global model. Each client uses the global model to make failure prediction. For each round of centralized learning, a central server trains the model through 40 epochs and each client uses this model to predict failures. Compared with federated learning model which is trained on local training dataset, the centralized training model is trained on an aggregated training dataset which consist of local datasets from all clients. We tested the centralized model and federated learning model using the same test dataset. For each round of local learning, each client trains a local model using local dataset through 40 epochs and uses the local model to make detection. 

The measurement results are shown in Fig.~\ref{lr-accuracy} and Fig.~\ref{nn-accuracy}. The x-axis represents the number of training round. The y-axis means the percentage of correct detection (both true positives and true negatives). In Fig.~\ref{lr-accuracy} and Fig.~\ref{nn-accuracy}, each sub-graph includes 4 lines, which shows the accuracy of model using different learning methods and tested on the corresponding client datasets. For example, in Fig.~\ref{lr-accuracy} \emph{Dataset \#1}, \emph{CDW\_Federated\_LR} represents the accuracy of the \emph{LR} model, which was trained using the proposed centroid distance weighted federated learning method with CDW\_FedAvg and tested on client \#1 dataset. Overall,  Fig.~\ref{lr-accuracy} and Fig.~\ref{nn-accuracy} show that the proposed CDW federated learning and standard federated learning can achieve satisfactory accuracy as local learning and centralized learning in most of the times. The accuracy of the models tested on clients differs from each other because the data distribution of the respective clients are different. Specifically, for both LR and NN, the accuracy of both the proposed CDW federated learning and standard federated learning tested on client \#3 is the highest, while the accuracy of federated learning performed on client \#4 is the lowest. Further, we observe that NN performs better than LR in all types of learning. In addition, the proposed CDW federated learning performs better than the standard federated learning for LR, while the two federated learning algorithms perform similar for NN.

In addition to accuracy, we also measured three other metrics in the experiments, including precision, recall, and F1-score. Precision means the proportion of truly classified positive instances to the total classified positive instances, while recall means the ratio of truly classified positive instances to the total number of practical positive instances. F1-Score is the harmonic mean of precision and recall. The formal definitions of Precision, Recall and F1-Score are shown in Formula (2)-(4) respectively. We measured these three metrics of LR and NN using both federated learning and centralized learning. The results are illustrated in Fig.~\ref{lr-metrics} and Fig.~\ref{nn-metrics}. The x-axis includes four metrics, while the y-axis is the the value of three measured metrics. The results show that the proposed CDW federated learning method and the standard federated learning method performed best on the dataset of client \#3, which is similar to the results of accuracy tests. Compared with the standard federated learning method, the proposed CDW federated learning method shows an improvement in Fig.~\ref{lr-metrics}, while the improvement is slight in Fig.~\ref{nn-metrics}.

\begin{equation}
Accuracy=\frac{TP+TN}{TP+TN+FP+FN}
\end{equation}
\begin{equation}
Precision=\frac{TP}{TP+FP}
\end{equation}
\begin{equation}
Recall=\frac{TP}{TP+FN}
\end{equation}
\begin{equation}
    F1 \verb|-| Score=\frac{2*Precision*Recall}{Precision+Recall}
\end{equation}

\emph{TP, TN, FP, FN} are the abbreviations of \emph{True Positive, True Negative, False Positive, False Negative}  respectively. 
~\\

\begin{table}[]
\small
\begin{tabular}{*{4}{r}}
\toprule
 Data Size & 10$^3$  & 10$^6$ & 10$^9$  \\
\midrule
 Centralized\_LR  & 10$^3$ & 10$^6$ & 10$^9$ \\
 Fed\_LR(1) & 28800 & 28800 & 28800 \\
 Fed\_LR(2) & 57600 & 57600 & 57600 \\
 Fed\_LR(3) & 86400 & 86400 & 86400 \\
 Fed\_LR(4) & 115200 & 115200 & 115200 \\
\bottomrule
\end{tabular}
\centering\caption{Communication overhead.}
\label{communication-overhead}
\end{table}

\textit{Test 4: Measuring the communication overhead}\par
We measured the communication overhead of LR. The results are shown in Table~\ref{communication-overhead}, where Fed\_LR(N) means N clients are chosen randomly to train the model in each round. In our implementation, the size of the \emph{LR} model is 144 bytes and the number of training rounds is set to 100. For instance, in Fed\_LR(3), the total communication overhead is 144*3*100*2=86400 bytes, where \emph{*2} means the models are gathered from every chosen client and then the global model is sent back to the clients after aggregating the client model updates. The table shows the overhead of centralized learning is much more than that of federated learning, because all data need to be sent to the central server from the client servers in centralized learning, while only model updates are sent to the central server in federated learning. Therefore, the communication overhead of the federated learning is irrelevant to the size of the data. When the size of dataset is extremely large, the communication overhead can be reduced significantly.

\section{Conclusion}
\label{conclusion}

This paper presents a blockchain-based federated learning approach for IIoT device failure detection. In the design, a central server coordinates client servers to train a shared global model for detection, with raw data stored locally. The architecture adopts blockchain to enable verifiable integrity of client data and incentives for client contribution. To reduce the impact of data heterogeneity, a novel centroid distance weighted federated averaging (CDW\_FedAvg) algorithm is proposed based on the distance between positive class and negative class of each client dataset.

A proof-of-concept prototype is implemented using the proposed architecture in a real-world use case. We evaluate the approach in terms of feasibility, detection accuracy, and performance. The evaluation results show that the proposed approach achieve all the proposed objectives. 

Although most of the federated learning modules are placed on client servers in this paper, we plan to adapt the design by moving all the modules from client servers to client devices with more powerful computing and storage capabilities in our future work. Also, we plan to explore how to increase the trustworthiness of client devices for the aggregation process in further study.



%

\section{Acknowledgement}
\label{Acknowledgement}

This research is supported by the National Natural Science Foundation of China (grant  62072469), National Key R\&D Program (2018YFE0116700), the Shandong Provincial Natural Science Foundation (ZR2019MF049, Parallel Data Driven Fault Prediction under Online-Offline Combined Cloud Computing Environment), and the Fundamental Research Funds for the Central Universities (2015020031).


\end{document}